# Dynamic film thickness measurement in a rolling bearing using numerical elastohydrodynamic-acoustic modelling to interpret reflected ultrasound data

Pan Dou[a, b], Yayu Li[a], Suhaib Ardah[b], Tonghai Wu[a]✉, Min Yu[b], Tom Reddyhoff[b], Yaguo Lei[a], Daniele Dini[b]

[a] Key Laboratory of Education Ministry for Modern Design and Rotor-Bearing System, Xi'an Jiaotong University, Xi'an, Shaanxi 710049, PR China

[b] Department of Mechanical Engineering, Imperial College London, SW7 2AZ, London, United Kingdom

✉Corresponding author: tonghai.wu@mail.xjtu.edu.cn

**Abstract**

*The thickness of the lubricating film plays a vital role in the operational efficiency and reliability of rolling bearings. Ultrasonic reflection techniques offer a promising non-invasive approach for in situ evaluation of lubricant films. However, accurately identifying the central film thickness remains challenging due to several complex factors, including dynamic fluctuations, localized elastic deformation, cavitation effects, and variations in oil supply. This study presents a comprehensive theoretical and numerical framework to elucidate the influence of these factors on ultrasonic wave propagation in lubricated contacts. Numerical simulations considering elastohydrodynamic lubrication (EHL) regime and cavitation-induced effects are carried out to obtain the surface deformation profiles and the cavitation regions. Subsequently, high-fidelity acoustic simulations are conducted to interpret reflected ultrasound data. As main results, the EHL leads to a "double-peak with central valley" pattern in the reflection coefficient distribution. While the cavitation causes the central valley to shift toward the inlet region and increases the reflection coefficient. Accordingly, the central film thickness is extracted from the distribution of the reflection coefficient under different operating conditions. Experimental validation using both glass–oil–steel and steel–oil–steel bearing setups confirms the effectiveness of the proposed method. The high-resolution fluorescence measurement adopted in the glass–oil–steel configuration validates the simulation of the reflection coefficient distribution. Furthermore, the theoretical EHL calculations are employed with the steel–oil–steel configuration for validation of the measurement accuracy of central oil film thickness.*

# 1. Introduction

Rolling bearings, serving as the "joints" of machines, possess advantages such as low friction coefficient, high radial load capacity, and high rotational accuracy, making them the core supporting components of large rotating equipment. Particularly, the operational reliability of the bearings is highly reliant on the lubricant film between the rollers and the inner and outer rings. Such oil film merely exists in micro-scale contact regions and supports all loads, therefore is extremely small and thin. These ultra-thin films is fragile to extreme conditions and is often ruptured to produce solid wear in roller bearings. Correspondingly, the lubricant film thickness monitoring can be the preliminary index for most failures of a rolling bearing. This provides a reliable reference for ensuring the operational stability of rotating equipment[1][2]. Many researchers are therefore motivated to inspect the oil film, however, it remains a significant challenge considering the complexities, including micro-scale, complex geometry, cavitation and dynamic deformations etc.

Up till now, different measurement techniques such as electrical, optical, and ultrasonic have been proposed for lubricant film thickness measurement. Although electrical and optical methods can achieve measurements of the lubricant film thickness under the EHL condition, they require strict electromagnetic shielding or the demand for at least one side of the friction pair to be transparent, thus limiting their applicability to laboratory environments[3][4]. The linear propagation and strong penetration of ultrasound make it the most promising technique for non-destructive, non-invasive measurement techniques of film thickness in industry settings[5]. Advances in acoustic measurement models have enabled the successful application of ultra-reflection-based technology in systems such as cylinder liner-piston ring systems, and fuel pumps [6][7]. However, its adoption in rolling bearings remains limited due to challenges in achieving sufficient spatial resolution for measurements.

To achieve the high-resolution measurement, various kinds of ultrasonic sensors are explored, including water-focused sensors, thin-film sensors, and piezoelectric ceramic sensors. However, water-focused sensors are bulky and difficult to install, while thin-film sensors involve complex fabrication processes, making both types challenging to apply in practical engineering scenarios [8][9]. In contrast, Clarke BP et al. proposed a promising approach by directly attaching adhesive piezoelectric ceramic sensors to the outer surface of roller bearings [10]. Due to the ability to cut and trim piezoelectric ceramic sensors to smaller widths, combined with their ease of installation, they have already been applied in some real-world engineering cases, such as large-scale rolling bearings [11]. Nevertheless, the width of the contact area to be measured remains larger than the width of the sensor itself.

To uncover the relationship between the central film thickness reflection coefficient and the overall sensor reflection coefficient from measurements—thereby capturing the coupled effects of multiple physical phenomena—relevant studies have made successful attempts using physics-based ray models[9], simulation-based acoustic models[12], and data-driven system identification methods[14]. This is primarily due to two factors: first, while the equivalent deformation assumption is widely adopted in elastohydrodynamic lubrication (EHL) calculations for its simplicity and computational convenience, it does not accurately capture the real profile geometry, resulting in discrepancies in ultrasonic wave scattering [9]. Second, the length of inlet region and the cavitation are commonly observed in rolling bearings, and the substantial difference between the reflection coefficients at steel–air and steel–oil film interfaces significantly alters the ultrasonic reflection behavior, introducing additional spurious reflections at the oil film–air boundary [15]. The coupling of these two effects leads to deviations in the correlation between the reflection coefficient of central film thickness ($R_c$) and the overall sensor ($R_{sc}$), which ultimately leads to measurement errors.

To address this issue, this paper adopts numerical elastohydrodynamic-acoustic modelling to interpret reflected ultrasound data under complex EHL conditions. First, EHL numerical simulation considering cavitation will be established to obtain critical information of line-contact lubrication. Afterward, a high-fidelity acoustic simulation can be accurately carried out to investigate the influence mechanisms of EHL, cavitation, inlet region length, speed and load on ultrasonic reflection. This allows the central film thickness to be determined from the reflection coefficient distribution under varying operating conditions. Finally, experimental validation is conducted using a roller bearing test rig. Simulation results are compared with fluorescence-based measurements, and the accuracy of the central film thickness is validated against theoretical EHL predictions. An overview of the proposed methodology is illustrated in Figure 1.

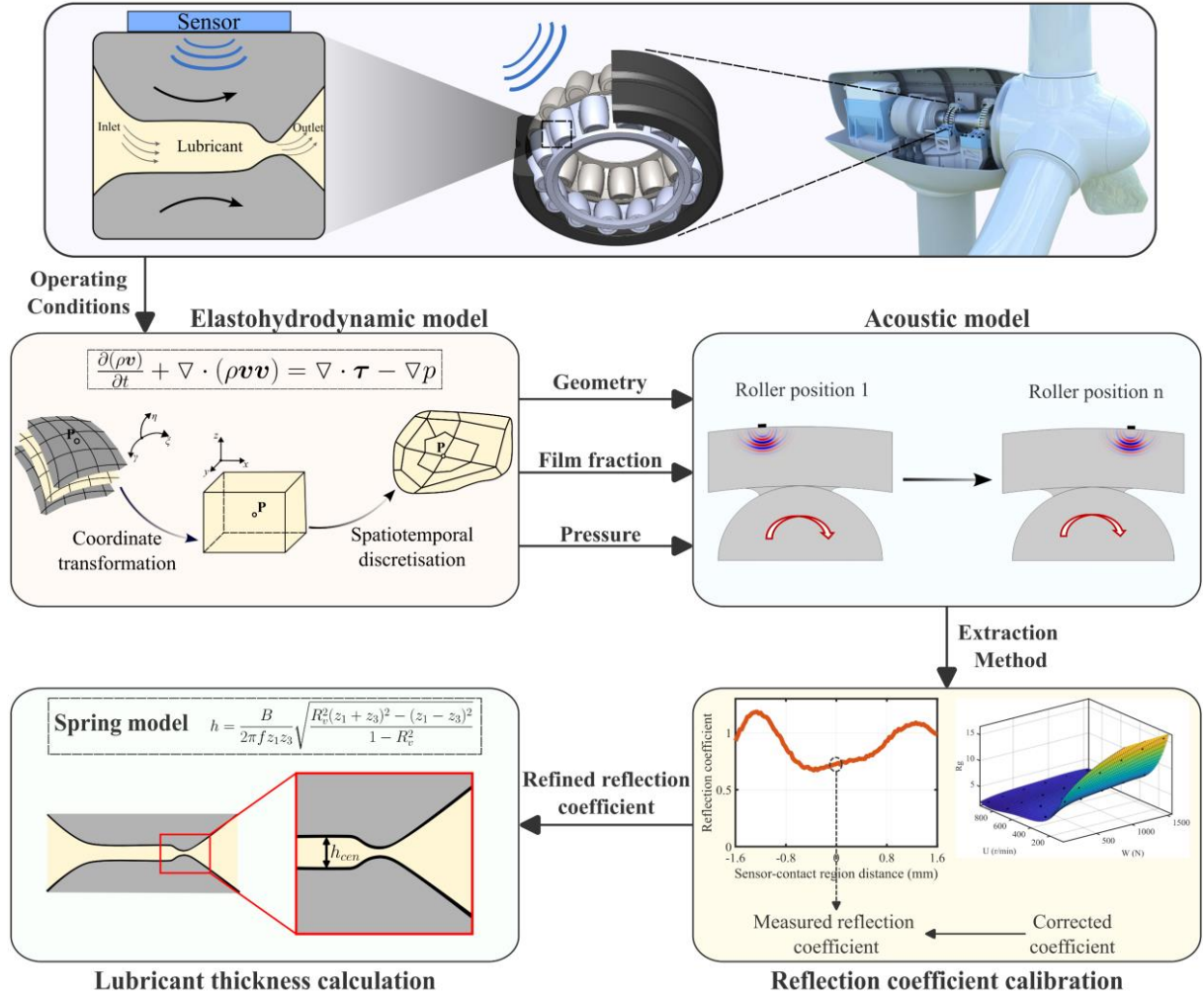

**Figure 1** Overview of the numerical elastohydrodynamic-acoustic modeling approach for interpreting ultrasonic reflections and measuring dynamic film thickness in a rolling bearing.

## 2. Methods

An EHL simulation is first performed to obtain detailed information on the deformed contact profile, oil film thickness distribution, position of cavitated regions, and hydrodynamic pressure field (Section 2.1). Subsequently, high-fidelity acoustic simulations are conducted to analyze the effects of outlet and inlet region lengths, load, and rotational speed on ultrasonic reflection behavior (Section 2.2). Based on these results, a method is proposed for determining the central film thickness using ultrasonic reflections (Section 2.3).

### 2.1 EHL Simulation considering cavitation

Elastohydrodynamic lubrication typically occurs in rolling element bearings operating under high-load and high-speed conditions. The distribution of the lubricant film is governed by a complex interplay of factors, most notably the hydrodynamic pressure generated in the confined film and the pressure-induced elastic deformation of the contacting surfaces. The lubricant film within a rooling or sliding contact is generally segmented into three distinct regions: inlet zone, Hertzian contact zone, and outlet zone. In the Hertzian region, located at the center of contact, the oil film thickness remains relatively uniform until it sharply decreases near the outlet. In the inlet region, the geometry and length of the lubricant film significantly affect pressure buildup; a longer inlet enhances the lubrication stability and reduces the risk of lubrication failure. In the outlet region, rapid pressure drop may cause cavitation, deteriorating lubrication performance and accelerating wear. To accurately determine the film thickness and pressure distribution, a cavitation-inclusive EHL simulation (Figure 2) is conducted.

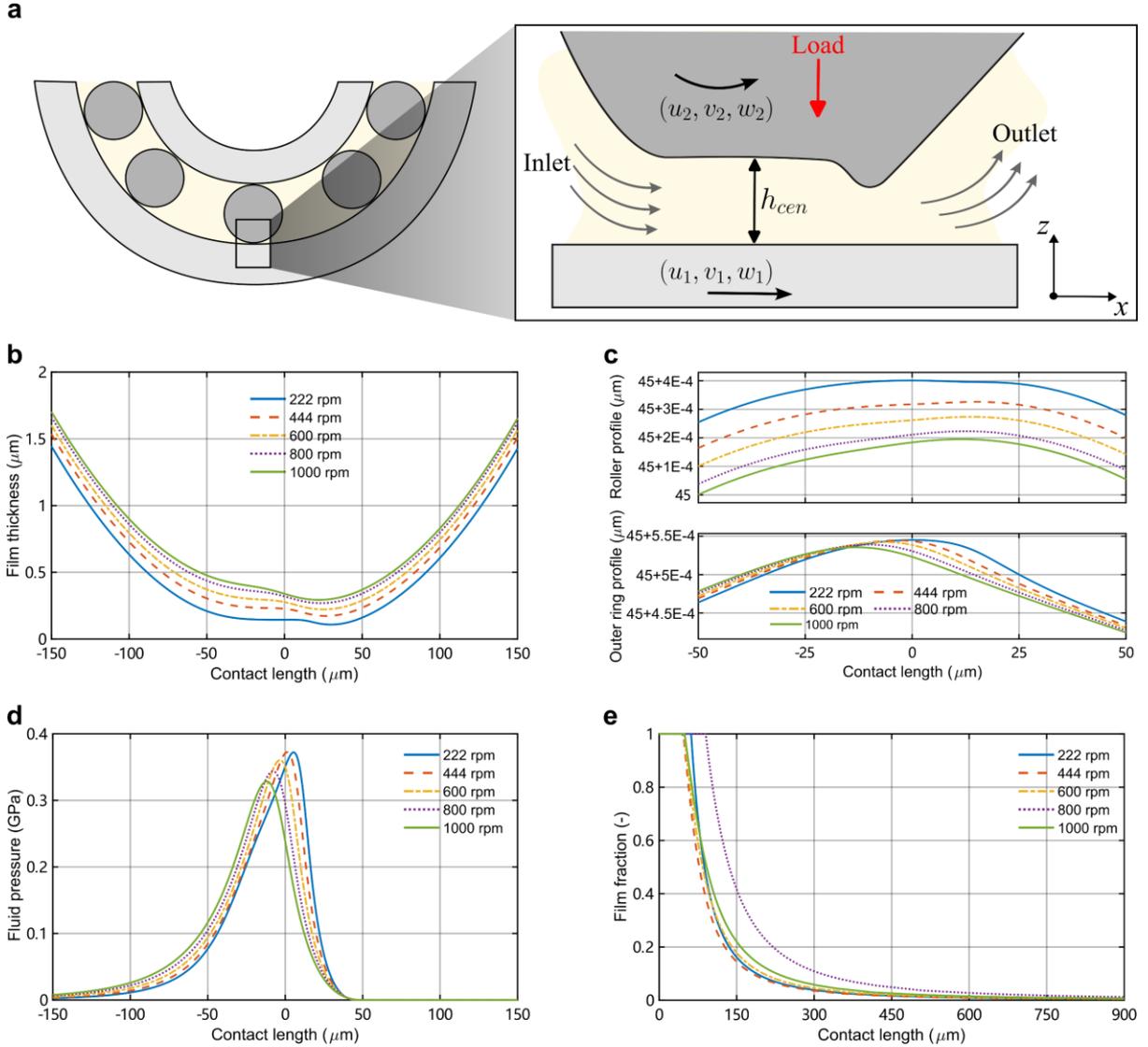

**Figure 2 EHL numerical simulation under line-contact lubrication conditions. a**, Schematic illustration of fluid film thickness and pressure distribution in an elastohydrodynamic lubrication (EHL) contact. **b**, Equivalent deformed profile of the contact based on EHL numerical calculation. **c**, Actual deformed profile showing elastic deformation under load. **d**, Pressure distribution within the lubricated contact region. **e**, Film fraction distribution indicating the presence of full film and cavitation regions.

The distribution of the hydrodynamic pressure and lubricant film content within the lubricated contact is described by the $p - \theta$ generalized Reynolds equation (GRE), which accounts for the combined effects of surface kinematics, film thickness variations $h(x, y, t)$, and spatiotemporal changes in lubricant properties such as viscosity $\eta(x, y, t)$ and density $\rho(x, y, t)$. Given the surface contact kinematics defined by the velocity fields $\boldsymbol{v}_1 = [u_1 \quad v_1]^T$ and $\boldsymbol{v}_2 =$ $[u_2 \quad v_2]^T$ along the x- and y-directions corresponding to the lower and upper surfaces, respectively and as per Figure 2, the modified $p - \theta$ generalized Reynolds equation is expressed in its conservative form as follows [14]:

$$\frac{\partial}{\partial x}\left(\epsilon \frac{\partial p}{\partial x}\right) + \frac{\partial}{\partial y}\left(\epsilon \frac{\partial p}{\partial y}\right) = \frac{\partial(\theta \rho_x^*)}{\partial x} + \frac{\partial(\theta \rho_y^*)}{\partial y} + \frac{\partial(\theta \rho_e)}{\partial t} \quad (1)$$

where $p$ denotes the local hydrodynamic pressure, and $\theta$ is the film fraction representing the liquid volume present in the biphasic mixture caused by cavitation. The corresponding coefficients and integrals are defined as follows:

$$\epsilon = \frac{\eta_e}{\eta_e'}\rho' - \rho'',$$

$$\rho_x^* = \rho_e u_1 + \eta_e \rho'(u_2 - u_1),$$

$$\rho_y^* = \rho_e v_1 + \eta_e \rho'(v_2 - v_1),$$

$$\rho_e = \int_{z_1}^{z_2} \rho \, dz,$$

$$\frac{1}{\eta_e} = \int_{z_1}^{z_2} \frac{dz}{\eta},$$

$$\frac{1}{\eta_e'} = \int_{z_1}^{z_2} \frac{z \, dz}{\eta},$$

$$\rho' = \int_{z_1}^{z_2} \rho \int_{z_1}^{z} \frac{dz'}{\eta} \, dz,$$

$$\rho'' = \int_{z_1}^{z_2} \rho \int_{z_1}^{z} \frac{z' dz'}{\eta} \, dz.$$

To enforce the mass conservation principles within the lubricated domain, this study employs the Elrod–Adams $p - \theta$ cavitation algorithm [16], which incorporates the Jakobsson–Floberg–Olsson (JFO) conditions as a foundation for defining cavitation boundaries. The associated complementary boundary conditions are specified as follows:

$$(p - p_{cav})(1 - \theta) = 0 \rightarrow \begin{cases} p > p_{cav} \rightarrow \theta = 1, & \text{pressured zones} \\ p = p_{cav} \rightarrow 0 \leq \theta < 1, & \text{cavitated zones} \end{cases} \quad (2)$$

where $p_{cav}$ represents the vapour-phase cavitation pressure corresponding to the pressure within the cavitation region. The parameter $\theta$ denotes the cavitation occupancy, indicating the local fraction of liquid lubricant. A value of $\theta = 1$ signifies a fully liquid-filled region, while $0 \leq \theta < 1$ corresponds to a cavitated region.

In line-contact lubrication, the characteristic dimensions and curvature radii of the contacting bodies are much larger than the width of the contact zone. This geometric setup permits the contact to be modeled under plane strain conditions, which are mathematically equivalent to a flat elastic medium experiencing a distributed pressure. The resulting film thickness distribution $h(x)$ along the one-dimensional contact can be expressed using [17]:

$$h(x) = h_0 + \frac{x^2}{2R'} + v(x), \tag{3}$$

where $h_0$ denotes the initial film thickness in the absence of elastic deformation, and $R'$ is the equivalent curvature radius of the contacting surfaces. The elastic surface deformation $v$, induced by a pressure distribution $p(s)$ acting on the line region $\Omega$, is calculated using the Flamant solution, as expressed below:

$$v(x) = -\frac{2}{\pi E'} \int_\Omega p(s) \ln(x-s)^2 \, ds. \tag{4}$$

The EHL model adopted in this study follows the computational framework presented in [17], where the governing equations are discretized through the Finite Volume Methods (FVMs). The nonlinearities exhibited by the system of governing equations are addressed using the semi-system approach integrated wihtin the iterative solution. The simulation framework employs the Point Gauss–Seidel method enhanced by the Aitken acceleration, collectively referred to as the Point Gauss-Seidel Method with Aitken Acceleration (PGMA). The PGMA scheme enhances the convergence stability of traditional partitioned solvers by adaptively adjusting the scalar relaxation factor within each iterative step, guided by the residuals computed from both the current and preceding iterative steps. For comprehensive details on the implementation of both the semi-system strategy and the PGMA technique, including their contributions to accelerating EHL simulation convergence and revamping numerical robustness under heavy load conditions, the reader is referred to [17]-[19]. Table 1 summarizes the key parameters of the roller and outer ring utilized in this study.

Table 1: Parameters for EHL simulation of the roller and ring contact

| Parameter | Roller | Outer ring |
|---|---|---|
| Radius ($R$/mm) | 6.25 | 45 (Inner radius) |
| Material | Steel | |

| Material elastic modulus (E/Gpa) | 210 |
| Poisson's ratio | 0.293 |
| Lubricant | PAO 4 |

The interaction between the roller and the outer ring is often idealized as a simplified contact model comprising an equivalent cylindrical body against a perfectly rigid foundation. However, this simplification deviates from the true geometrical configuration of the rolling element-raceway contact. Most importantly, the propagation behavior of ultrasonic waves are sensitive to geometric details, and equivalent shapes may fail to capture the wave behavior accurately. Therefore, it is vital to reconstruct the actual geometric deformation profiles based on accurate geometrical relationships. Based on the equivalent deformation field, the individual deformation contributions from the roller and the inner surface of the inner ring can be calculated as follows:

$$v_r(x) = v(x)\frac{(1-\mu^2)E'}{2E_1}, \tag{5}$$

$$v_{ri}(x) = v(x)\frac{(1-\mu^2)E'}{2E_2}, \tag{6}$$

where $v(x)$ is the equivalent deformation, while $v_r(x)$ and $v_{ri}(x)$ denote the deformation fields of the roller and the inner surface of the inner ring, respectively.

Leveraging the deformation of the inner surface of the inner ring and its initial geometric profile, the true geometric profile of the inner ring, denoted as $d_{ri}(x)$, can be calculated using:

$$d_{ri}(x) = R_{ri} - \sqrt{R_{ri}^2 - x^2} + v_{ri}(x), \tag{7}$$

where $R_{ri}$ is the inner radius of the inner ring.

On the other hand, the geometric profile of the outer surface of the inner ring, denoted by $d_{ri}(x)$, is determined by its initial geometric profile as follows:

$$d_{ro}(x) = R_{ro} - \sqrt{R_{ro}^2 - x^2}, \tag{8}$$

where $R_{ro}$ is the outer radius of the inner ring.

Based on the deformation of the inner surface of the roller and its initial geometric profile, the true geometric profile of the roller, denoted by $d_r(x)$, can be evaluated using:

$$d_r(x) = R_{ro} - \sqrt{R_{ro}^2 - x^2} + v_{ro}(x), \tag{9}$$

where $R_{ro}$ is the inner radius of the inner ring.

Figures 2b to 2e illustrate the the simulation results obtained under a load of 230 N at various rotational speeds, including the associated hydrodynamic pressure and film fraction profiles. Significant discrepancies exist between the equivalent and actual deformed profiles, as depicted in Figures 2b and 2c, which lead to notable variations in acoustic wave propagation, reflection, and transmission. Furthermore, the pressure distribution shown in Figure 2d is utilized to calculate the oil's density and sound velocity variations along the distance in the ultrasonic simulation. The film fraction, illustrated in Figure 2e and derived from Equation (2), indicates the presence of cavitation. The initial location of the cavitation region can be used to determine the outlet region length in the following analysis.

### 2.2 Acoustic simulation of ultrasonic reflection under EHL conditions

A high-fidelity acoustic simulation model is developed to investigate the influence of EHL conditions on ultrasonic reflection, as shown in Figure 3. The model simulates the propagation of ultrasonic waves in a rolling bearing based on COMSOL Multiphysics software. The specific steps to establish the finite element simulation model of the acoustic field are as follows:

1) Step 1: Geometric model
2) Step 2: Material parameters
3) Step 3: Boundary conditions
4) Step 4: Excitation signal
5) Step 5: Mesh generation
6) Step 6: Solving

The geometric model, including the oil film thickness distribution and the solid bodies' geometric profiles, can be transplanted from the results in Figure 2. For facilitation, the relative moment between the roller-ring is simplified as varied positions of the ultrasonic sensor referring to the fixed roller. This does not change the essence of the roller bearing. The sensor moving direction and the oil film distribution are exhibited in Figure 3a. An acoustic simulation is conducted based on four variables: the inlet region length, the outlet region length, rotational speed, and load. A detailed explanation of steps 2–6 is provided in [12].

For each sensor position, the time-domain echo signals are transformed into the frequency domain using Fast Fourier Transform (FFT), yielding the corresponding reflected frequency signals. The reflection coefficient is then calculated as the amplitude ratio between the echo signal and a reference signal obtained by simulating the oil film as an air medium. A controlled variable method is employed to investigate the effect of each factor on the ultrasonic reflection coefficient:

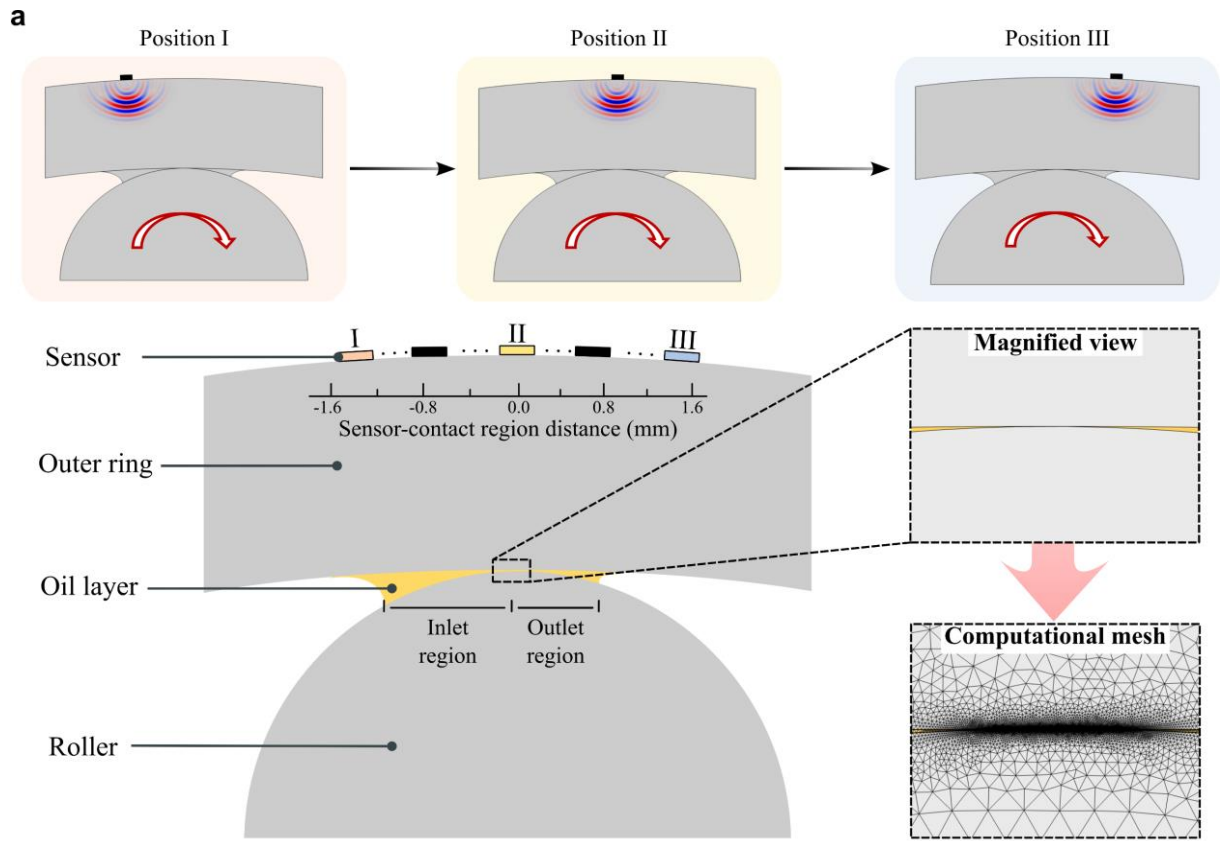

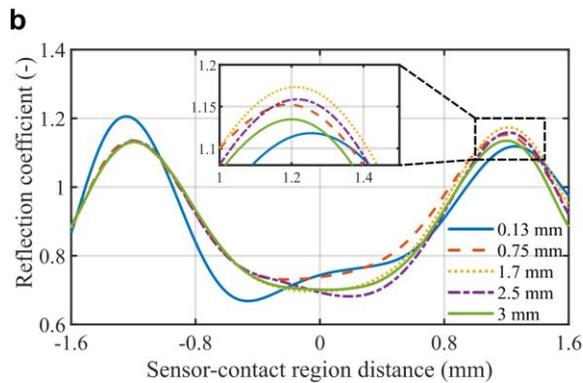
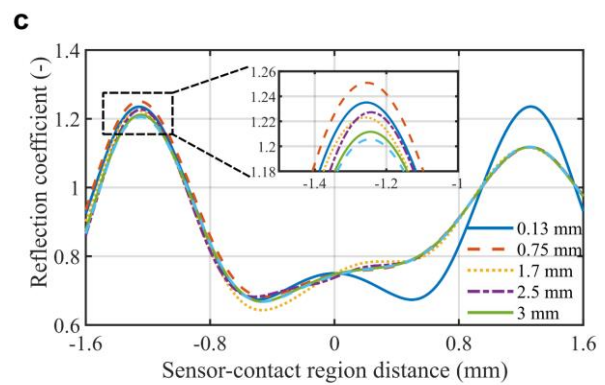
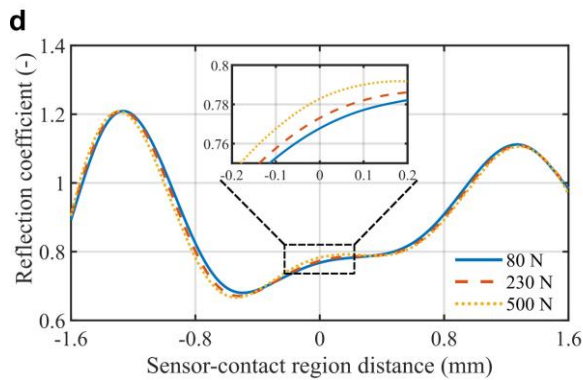
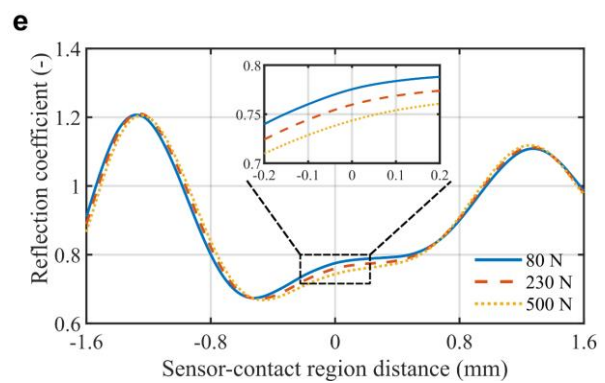

**Figure 3 | Acoustic simulation of ultrasonic reflection under EHL conditions. a,** Geometric model of the roller–inner ring contact. **b,** Effect of outlet region length on reflection coefficient, assuming fixed values for other parameters. **c,** Effect of inlet region length on reflection coefficient, assuming fixed values for other parameters. **d,** Effect of rotational speed on reflection coefficient, assuming fixed values for other parameters. **e,** Effect of load on reflection coefficient, assuming fixed values for other parameters.

(1) Effect of outlet region length

Figure 3b illustrates how the reflection coefficient changes with sensor-to-contact distance under different outlet lengths of the oil film, with the inlet region length fixed at 3 mm, the rotational speed set to 400 rpm, and the load set to 230 N. It can be observed that when both the outlet and inlet regions are fully lubricated(i.e., equal to 3 mm), the EHL results in a characteristic "double-peak–single-valley" pattern in the reflection coefficient distribution. The two side peaks are symmetrical, while the central valley is relatively pronounced.

The peak value within the outlet region increases and then decreases as the outlet region length is reduced, and it is consistently higher than the peak value in the inlet region. The peak value in the inlet region remains relatively unchanged as the outlet region length decreases, but it suddenly increases when the outlet length becomes comparable to the sensor width (around 0.3 mm). Additionally, near the contact area (approximately -0.5 mm to 0.5 mm), as the outlet length decreases, A slight "collapse" occurs toward the right of the central region, causing the valley to shift toward the outlet side. When the outlet region gradually decreased to 0.13 mm, which is the cavitation initial position determined by the EHL simulation, the "collapse" occurs toward the right of the central region, causing the valley to shift toward the outlet side. What's more, the reflection coefficient corresponding to the central oil film thickness increases as the outlet region length is reduced.

(2) Effect of inlet region length

Figure 3c illustrates how the ultrasonic response changes relative to sensor-to-contact center distance under different inlet lengths of the oil film, with the outlet region length set to 0.13 mm and a rotational speed of 400 rpm. As the inlet region length decreases, the peak value within inlet region gradually increases, while the outlet peak on the right side changes minimally. When the inlet region length approaches the sensor width, the peak value within inlet region gradually decreases, and at the same time, the peak value in the outlet region on the right side begins to increase. Additionally, it can be observed that the variation of inlet region length has a relatively minor impact on the reflection coefficient at the center of the contact area.

(3) Effect of rotational speed

Figure 3d illustrates how the ultrasonic response changes relative to sensor-to-contact center distance under different rotational speeds, with the load set to 230 N. It can be observed that the reflection coefficient near the contact region (approximately -0.5 mm to 0.5 mm) gradually decreases as the rotational speed increases.

(4) Effect of load

Figure 3e illustrates how the ultrasonic response changes relative to sensor-to-contact center distance under different loads, with the rotational speed set to 400 rpm. The reflection coefficient near the contact region (approximately -0.5 mm to 0.5 mm) gradually increases as the load increases.

The following conclusions can be drawn from above analysis:

(1) In the absence of cavitation, the elastohydrodynamic (EHD) deformation profile results in a characteristic "double-peak–single-valley" pattern in the reflection coefficient distribution. The two side peaks are symmetrical, while the central valley is relatively pronounced.

(2) The presence of cavitation increases reflection coefficients corresponding to central oil film thickness. A slight "collapse" occurs toward the left of the central region, causing the valley to shift toward the inlet side.

(3) The decrease in inlet region length causes the side peak in the inlet region to first increase and then decrease, but has a minimal impact on $R_c$.

(4) Under different loading and rotational speed conditions, the general trend of the reflection coefficient distribution curves remains consistent. However, due to the presence of cavitation, $R_c$ increases with increasing rotational speed and load.

### 2.3 Central oil film thickness extraction

The peak values in both the outlet and inlet regions are approximately symmetrical under varying inlet length, outlet length, load, and rotational speed. Therefore, the symmetric center between the two peaks serves to locate the extraction point for accurately obtaining $R_c$. Additionally, the inlet region length has a minimal impact on this reflection coefficient. The deformation profile along with cavitation initial location can be determined through EHL simulation. An increase in applied load and rotational velocity leads toward a rise in measured reflection coefficient corresponding to the central lubricant film. Thus, the extraction method for the central film thickness can be realized through EHL and acoustic simulations. The procedure is outlined below:

Step 1: Perform EHL simulations to calculate the deformation profile, contact pressure distribution, and cavitation initiation location of the roller-inner ring contact area under different operating conditions (e.g., load, speed).

Step 2: Based on acoustic simulations, establish an acoustic model for the roller-inner ring system to obtain the reflection coefficient $R_{fz(centre)}$ in the central region and the overall sensor reflection coefficient $R_{fz}$ under various operating conditions.

Step 3: In order to define the mapping between these two reflection coefficients, an adjustment factor $R_g$ is introduced:

$$R_g = \frac{R_{fz(centre)}}{R_{fz}} \tag{10}$$

Further, a polynomial fitting method is used to establish the relationship between the adjustment factor and operating parameters. Assuming the fitting model is an nnn-th order polynomial, the equation is expressed as:

$$R_g = \sum_{i=0}^{n} a_i U^i L^{n-i} \tag{11}$$

where $a_i$ are the fitting coefficients, $U$ represents the rotational speed, and $L$ represents the load. Using the least squares method or other fitting techniques, the coefficients aia_iai are determined, and the most suitable polynomial order nnn is selected based on the fitting accuracy required for the actual data. This polynomial may include linear, quadratic, and interaction terms, with the degree nnn chosen based on the data fitting accuracy.

Step 5: In actual measurements, when the reflection coefficient $R_{sc}$ obtained from sensor data is known, together with bearing dimensions, load, and speed are known, the corresponding adjustment factor $R_g$ for the operating conditions can be determined from the polynomial fitting results. Subsequently, the reflection coefficient $R_c$ in the contact area can be calculated:

$$R_c = R_{sc} \cdot R_g \tag{12}$$

Step 6: Using the corrected reflection coefficient $R_v$, the average film thickness $h$ in the contact zone can be obtained using the spring model method[20]. The formula is as follows:

$$h = \frac{B}{2\pi f z_1 z_3} \sqrt{\frac{R_c^2 (z_1 + z_3)^2 - (z_1 - z_3)^2}{1 - R_c^2}} \tag{13}$$

where, $h$ represents the lubricant film thickness, and $f$ is the frequency of the sound wave, typically chosen as the sensor's center frequency. The term $z = \rho c$ denotes the acoustic impedance of the medium, with subscripts 1 and 3 representing the tribo-pair materials on either side of the lubricant film.

This algorithm, by combining EHL calculations with acoustic simulations, establishes the relationship between the ultrasonic reflection coefficient and film thickness, fully accounting for the real deformation profile, contact pressure distribution, and cavitation effects. Compared to the traditional spring model calculation, this method provides a more accurate measurement of the central film thickness in the contact area.

## 3. Experiments

To validate the influence of EHL deformation, outlet region length, and inlet region length on ultrasonic reflection, an experiment was first conducted on a rolling bearing test rig using a glass–oil–steel configuration—with a glass outer ring and steel rollers. The results were validated against fluorescence data to support the simulation-derived mechanism analysis. Subsequently, a steel–oil–steel configuration—with both the outer ring and rollers made of steel was employed to further validate the accuracy of the proposed central film thickness extraction method under realistic bearing conditions by comparing the measured results with theoretical EHL calculations.

### 3.1 Experiment with glass-oil-steel configuration

Figure 4 shows experimental setup and results for the glass–oil–steel bearing configuration used to verify ultrasonic reflection mechanisms under EHL conditions. The test rig employed a cylindrical roller bearing, model N208, with its outer ring replaced by a glass ring of the same size without a raceway but with a coating. Through an optical interferometer, the fluorescence method allowed for direct observation of the lubricant oil layer on the inner cylindrical surface of the glass outer ring. The fluorescence method measured oil film thickness by adding a fluorescent dye to the lubricant and illuminating it with ultraviolet light. The emitted fluorescence intensity exhibits a direct correlation with film thickness, allowing quantitative measurement through calibration[4].

Figure 4b illustrates the photograph for attaching the ultrasonic sensor to the glass outer ring of the bearing. The ultrasonic sensor used was a rectangular piezoelectric ceramic sensor, which had been encapsulated and wired for stability. To facilitate both optical observation and the attachment of the ultrasonic piezoelectric ceramic sensor, the measurement position of the ultrasonic sensor was located at the uppermost roller.

The specific experimental procedure was as follows: a servo driver was used to control the motor for precise speed control of the bearing's inner ring, with the rotational speed range in the experiment being 50-1000 r/min. A loading screw was used to apply radial loads in a vertically downward direction. Since the load had a minimal effect on lubricant film thickness, the maximum pressure at this measurement point was maintained at 0.18 GPa, the contact area width at 80 μm, and the room temperature at a constant 23°C during the experiment. The lubricant used in the experiment was model PAO4, with a dynamic viscosity of 0.027 Pa·s. During the experiment, a pipette was used to add a precise 1 ml of lubricant to the bottom of the inner cylindrical surface of the outer ring, ensuring the lubricant was evenly distributed across the surfaces of the outer ring, rollers, and inner ring by driving the bearing.

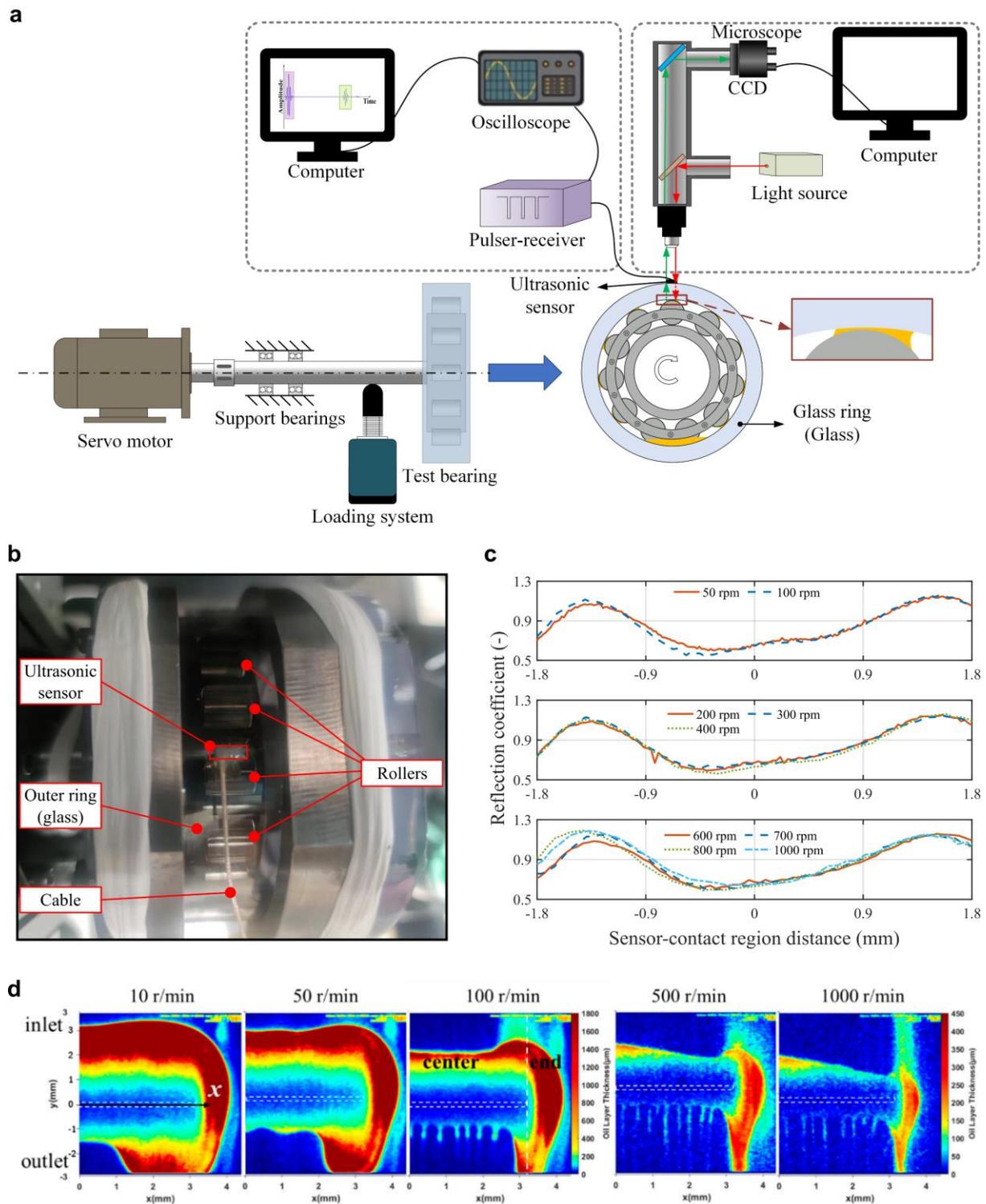

**Figure 4 Experimental setup and results for the glass–oil–steel bearing configuration used to verify ultrasonic reflection mechanisms under EHL conditions. a**, Schematic of the test rig setup integrating ultrasonic and optical fluorescence measurements. **b**, Installation of the ultrasonic sensor for in-situ measurement in the glass bearing setup. **c**, Measured reflection coefficients under different rotational speeds using PAO 4 oil. **d**, Optical fluorescence measurements under different rotational speeds using PAO 8 oil [2], used for comparison and validation of the ultrasonic reflection measurements.

Figure 4c shows the measurement results regarding the ultrasonic reflection coefficient during the passage when the roller moves at different speeds (50-1000 r/min) under the same load conditions (PAO 4 oil). From Figure 4c, it can be observed that between -1.5 mm and 1.5 mm, the overall shape exhibits a 'double-peak, single-valley' pattern. As the rotational speed increases, the inlet-region peak value (within the purple dashed box) remains unchanged. In contrast, the outlet-region peak amplitude (within the red dashed box) increases with rotational speed, particularly in the exit area. Additionally, it can be observed that the minimum value point consistently appears in the inlet region length.

Figure 4d presents the measurement results using the fluorescence method (PAO 8 oil). As the speed increases, the inlet region length indeed decreases, and cavitation appears in the outlet region. The position of the cavitation shows little variation with changes in rotational speed. Additionally, it can be observed that the inlet region length is always greater than the outlet region length. This is consistent with the changes observed in the measurement results for the reflection coefficient in Figure 4c.

### 3.2 Experiment with steel-oil-steel configuration

The rolling and sliding test rig was used to validate the ultrasonic measurement technique of the bearing with steel outer ring and steel rollers (Figure 5). The rolling and sliding test rig primarily consisted of: the test bearing, loading system, servo drive system, industrial control computer, and lubrication system. The operating principle was as follows: the loading system applied a load to the outer ring of the bearing while keeping the outer ring fixed in place. The industrial control computer controlled the servo motor, which rotated the bearing's inner ring at various speeds. Lubrication oil was supplied by a peristaltic pump sprayed onto the rollers and outer ring, ensuring lubrication of the contact area. The test bearing model was a standard bearing with a flanged outer ring, model NF211EM. In the experiment, the slip ratio was set to 0, and the rollers underwent pure rolling motion. A total of 12 operating conditions were set, where the load on the rollers at different measurement points was 86.25N, 143.75N, and 230N, and the inner ring speeds were 166.5 rpm, 222 rpm, 277.5 rpm, and 333 rpm, respectively.

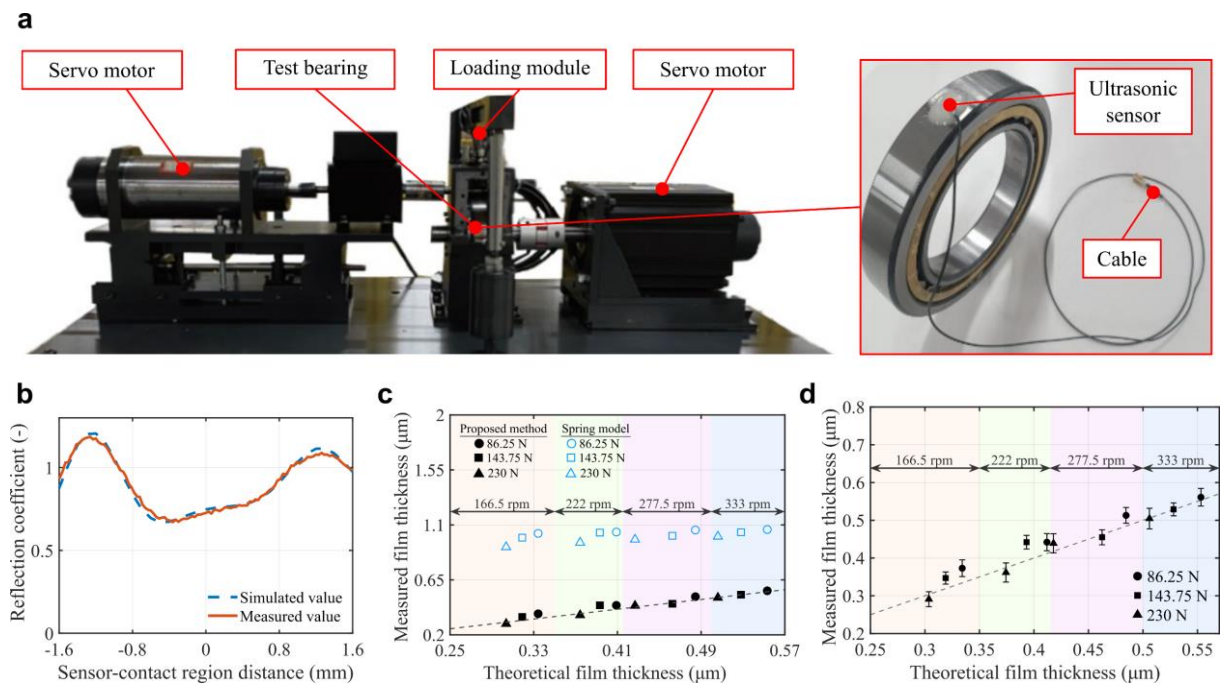

**Figure 5 | Experimental validation of central film thickness extraction using a steel–oil–steel bearing configuration. a**, Photograph of the bearing test rig with a steel outer ring and steel rollers. **b**, Comparison between simulated and experimentally measured reflection coefficient distributions at 230 N load and 333 rpm. **c**, Comparison of central film thickness measurement results obtained by different methods. **d**, Central film thickness values measured using the proposed method for 10 different rollers under identical operating conditions.

Figure 5b shows the comparison between the simulated and measured reflection coefficient distributions when the load is 230 N and the speed is 333 rpm. It can be observed that the simulated reflection coefficient distribution matches well with the measured results at the first peak in the inlet region and the first peak within the outlet region. The discrepancies arise from multiple sources, with the primary deviation likely due to the simplified and idealized setting of cavitation in the simulation. In reality, the cavitation shape is much more complex, featuring not only oil–gas transition regions but also exhibiting "fern cavities" or "oil strips."[22] These irregularities in the oil distribution can significantly affect ultrasonic wave reflections and introduce errors into the results.

Figure 5c shows central oil film thickness measurement results across various operating conditions. It is apparent that when the overall reflection coefficient of the sensor is directly used in the spring model formula to estimate the film thickness, the deviation relative to the EHL theoretical prediction is largest. In contrast, the proposed method produces oil film thickness measurements that closely align with EHL theoretical predictions under various operating conditions, with a maximum error of 12.7%. Therefore, the proposed method achieves higher measurement accuracy. Additionally, as shown in Figure 5d, fluctuations can be observed in the measurement results for different rollers under identical operating conditions. This could be due to signal interference and fluctuations in the measurement instruments, vibrations of the test rig, or inconsistencies in roller manufacturing.

## 4. Conclusion

Aiming at dynamic oil film thickness measurements in rolling bearings, this paper adopts numerical elastohydrodynamic-acoustic modelling to interpret reflected ultrasound data under complex EHL conditions. Key findings indicate that: (1) Without cavitation, the reflection coefficient distribution shows a typical "double-peak with central valley" shape, with nearly symmetrical side peaks and a relatively deep central valley. (2) Cavitation shifts the valley towards the inlet region side and increases $R_c$. From these observations, the symmetric center between the two peaks is used to determine the extraction point for $R_c$. The extract method of central film thickness by combining EHL and acoustic simulations is established under different operating conditions, enabling accurate extraction of the central film thickness.

Experimental validation through both glass–oil–steel and steel–oil–steel bearing setups confirms the effectiveness of the proposed method. The glass–oil–steel configuration, compared against fluorescence measurements, validates the simulation-based analysis of reflection coefficient distribution. The steel–oil–steel configuration further confirms the method by comparing the extracted central film thickness with theoretical EHL calculations. Compared to the traditional approache, the proposed method significantly reduces measurement errors, with a maximum error of 12.7%, demonstrating a substantial improvement in measurement accuracy.

## Acknowledgments

The authors gratefully acknowledge the financial support from the National Natural Science Foundation of China (Grant Nos. 52405597 and 52275126). S. Ardah and D. Dini would like to acknowledge the support received from the Engineering and Physical Sciences Research Council, United Kingdom (EPSRC) via the InFUSE Prosperity Partnership EP/V038044/1. D. Dini acknowledges the support of the Royal Academy of Engineering (RAEng) for the Shell/RAEng Research Chair in Complex Engineering Interfaces.